\pdfoutput=1
\documentclass[12pt,journal,twocolumn]{IEEEtran}
\usepackage[utf8x]{inputenc}
\usepackage{amsmath}
\usepackage{booktabs}
\usepackage{graphicx}
\ifCLASSOPTIONcompsoc
\usepackage[caption=false,font=normalsize,labelfont=sf,textfont=sf]{subfig}
\else
\usepackage[caption=false,font=footnotesize]{subfig}
\fi
\usepackage{url}
\usepackage{tikz}
\usepackage{textcomp}
\usepackage{hyperref} 

\newcommand{\fref}[1]{Fig.~\ref{#1}}

\graphicspath{
	{images/}
}
\IEEEoverridecommandlockouts

\newcommand\copyrighttext{%
	\footnotesize \textcopyright 2020 IEEE. Personal use of this material is permitted.
	Permission from IEEE must be obtained for all other uses, in any current or future
	media, including reprinting/republishing this material for advertising or promotional
	purposes, creating new collective works, for resale or redistribution to servers or
	lists, or reuse of any copyrighted component of this work in other works.
	DOI: \href{https://doi.org/10.1109/MCOM.001.2000722}{10.1109/MCOM.001.2000722}}
\newcommand\copyrightnotice{%
	\begin{tikzpicture}[remember picture,overlay]
		\node[anchor=north,xshift=0pt,yshift=-10pt] at (current page.north) {\fbox{\parbox{\dimexpr\textwidth-\fboxsep-\fboxrule\relax}{\copyrighttext}}};
	\end{tikzpicture}%
}

\title{Adding Indoor Capacity Without Fiber Backhaul:\\a mmWave Bridge Prototype}

\begin{document}
\newdimen\origiwspc%
\origiwspc=\fontdimen2\font%
\fontdimen2\font=0.85\origiwspc%
\linespread{0.98}


\bstctlcite{ShortCTL:BSTcontrol}
\author{
	\IEEEauthorblockN{
		Adrian Schumacher\textsuperscript{*}\textsuperscript{\#},
		Ruben Merz\textsuperscript{*} and
		Andreas Burg\textsuperscript{\#}}\\
	\IEEEauthorblockA{\textsuperscript{*}
		Swisscom (Switzerland) Ltd., CH-3050 Bern, Switzerland\\
		\{adrian.schumacher, ruben.merz\}@swisscom.com}\\
	\IEEEauthorblockA{\textsuperscript{\#}
		Telecommunications Circuits Laboratory, EPFL, CH-1015 Lausanne, Switzerland\\
		andreas.burg@epfl.ch\vspace{-0.5\baselineskip}}
\vspace{-\baselineskip}}
\maketitle
\thispagestyle{empty}
\copyrightnotice

\vspace{-\baselineskip}
\begin{abstract}
Today, a large portion of the mobile data traffic is consumed behind the shielding walls of buildings or in the Faraday cage of trains.
This renders cellular network coverage from outdoor cell sites difficult.
Indoor small cells and distributed antennas along train tracks are often considered as a solution, but the cost and the need for optical fiber backhaul are often prohibitive.
To alleviate this issue, we describe an out-of-band repeater that converts a sub-6\,GHz cell signal from a small cell installed at a cell tower to a mmWave frequency for the fronthaul to buildings or distributed antenna sites, where the signal is downconverted to the original frequency and emitted for example inside a building.
This concept does not require fiber deployment, provides backward compatibility to equipment already in use, and additional indoor capacity is gained while outdoor networks are offloaded.
The architecture and hardware prototype implementation are described, and measurements are reported to demonstrate the functionality and compatibility with commercial infrastructure and mobile terminals.
\end{abstract}


\section{Introduction}
\label{sec:introduction}
Because of the popularity of video streaming and social networking, mobile network data traffic has been growing in the last decade and this trend is predicted to continue \cite{EricssonAB_EricssonMobilityReport_2020}.
Thanks to the convenience and reliability, people access the Internet with their cellular devices also at home, in offices, and inside other buildings.
Statistics from studies such as \cite{abiresearch_inbuildingtraffic_2016} and our estimates on the Swisscom network show that around 80 percent of the mobile data traffic now originates or terminates indoors. Particular circumstances like the COVID-19 lockdowns in the first half of 2020 reinforce the indoor use \cite{EricssonAB_EricssonMobilityReport_2020}.
However, mobile network operators deploy most of their cellular sites outdoor.
To keep up with the demand for higher data capacity, wireless communication standards have introduced measures to improve the spectral efficiency, such as multiple input multiple output (MIMO) and means to reduce inter-cell interference.
Unfortunately, these measures alone are insufficient to meet the demand for higher capacity and more bandwidth is needed. As bandwidth in frequencies below 3\,GHz is scarce, the solution is to resort to higher frequencies.
With the introduction of the 5th generation (5G) of mobile cellular networks, wider bandwidths are becoming available at much higher frequencies than traditionally used, which are commonly referred to as millimeter-wave (mmWave) frequencies.
The disadvantage of using mmWave frequencies is a higher propagation attenuation, particularly when penetrating buildings \cite{ITUR_CompilationMeasurementDataRelating_2017}.
Commonly used low emissivity (Low-E) coated glass windows for thermal isolation also strongly attenuate radio frequency (RF) signals already at frequencies below 6\,GHz (sub-6\,GHz) \cite{Schumacher_3.5GHzCoverageAssessment_2019}.
Therefore, indoor users cannot immediately benefit from adding high-capacity cells in mmWave frequencies to existing cellular network outdoor sites.

The solution to provide large data capacities indoors is to deploy the wireless network also inside buildings using small cells (including various forms like picocells and femtocells).
Despite the advantages of small cells, such as reduced inter-cell interference, better uplink capacity due to lower path losses, and improved macro cell reliability \cite{Chandrasekhar_FemtocellNetworksSurvey_2008},
small cell infrastructure is often costly to deploy.
Specifically, meeting the stringent backhaul latency and data rate requirements ($>$1\,Gb/s) is difficult and expensive, especially when new optical fiber connections are required \cite{Xie_CostComparisonsBackhaulTransport_2018}.
Providing better mobile network capacity indoors can also be achieved with relays (decode-and-forward) or with repeaters (amplify-and-forward). However, these solutions do not add additional capacity indoors, but only bridge the capacity gap between indoor and outdoor links.

\begin{figure}
	\centering
	\includegraphics[width=\linewidth]{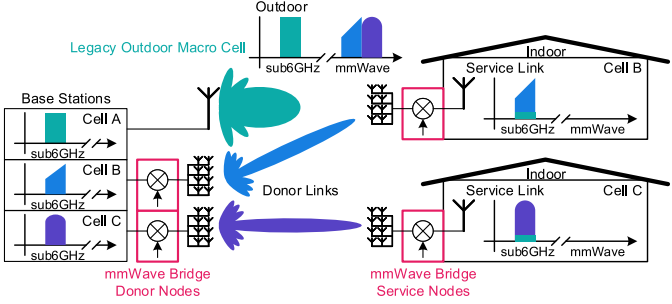}
	\vspace{-2em}
	\caption{Two sub-6\,GHz cells (B and C, same carrier frequency) are upconverted to mmWave, transmitted to a building, where they are downconverted and re-transmitted inside on their original carrier frequency. Indoor users are offloaded from the legacy outdoor cell A to either cell B or C.\label{fig:mmwb_concept_example}}
\end{figure}

To alleviate the issues described above, we propose a \emph{mmWave bridge} that provides sub-6\,GHz mobile network cells inside buildings or in areas with poor coverage, without the need for costly optical fiber links.
Instead of relying on optical fiber backhaul for an indoor small cell, the RF signals are transported to and from the buildings over mmWave frequencies.
Our mmWave bridge consists of a \emph{donor node} typically installed on an existing cell tower and cabled to a small cell base station (BS) that is also added; see \fref{fig:mmwb_concept_example} on the left side. The BS provides the donor cell at a sub-6\,GHz frequency, which is upconverted to a given mmWave frequency and transmitted to the mmWave bridge \emph{service node}, also known as the customer premise equipment (CPE). The CPE downconverts the cell signal to the original sub-6\,GHz frequency, which is then fed inside the building and transmitted to the user equipment (UE), as shown on the right in \fref{fig:mmwb_concept_example}.
The uplink operates in reverse direction.
In essence, this can be seen as an analog RF fronthaul over an out-of-band amplify-and-forward relay.

\subsection{Related Work}
Various canonical solutions for providing higher wireless data capacities indoor are described and compared in \cite{Schumacher_MmWaveBridgeConceptSolve_2020}.
With Release~16 of the 5G New Radio (NR) standard, integrated access and backhaul (IAB) \cite{Teyeb_IntegratedAccessBackhauledNetworks_2019} is introduced. IAB allows out-of-band relaying of sub-6\,GHz cells over mmWave frequencies. Because the IAB node acts itself as a BS, this solution is more complex than a mmWave bridge service node. Further, additional latency is introduced because of protocol processing in the IAB node.\looseness=-1

A low-cost amplify-and-forward relay node is described in \cite{MBreiling_UE-SideVirtualMIMOUsing_2014}, where the aim is to counteract limited spatial diversity in outdoor-to-indoor scenarios. To this end, spatially multiplexed massive MIMO signals from an outdoor donor cell in a sub-3\,GHz frequency band are translated to frequency multiplexed service signals at mmWave frequencies. Deployed in large numbers inside buildings, this concept promises high capacity gains thanks to a higher MIMO channel rank and multi-user MIMO. It relies on a low building entry loss and many uncorrelated MIMO paths. Unfortunately, such a system is not compatible with existing standard-compliant equipment: BS as well as UE need extensions to their radio access technology (RAT) protocol stacks and chipsets.\looseness=-1

A concept that translates sub-3\,GHz signals to mmWave frequencies and back is described in  \cite{Zhu_Millimeter-waveMicrowaveMIMORelays_2018}. The goal is also to increase the channel capacity by providing a high MIMO channel rank inside a building. At the donor cell, the baseband signal is time-space-coded, and the multiple MIMO layers are allocated on different frequency sub-bands resulting in a single input single output (SISO) signal with a larger bandwidth. The complete block of sub-bands is upconverted to mmWave frequencies and transmitted to the relay unit at a building, where the sub-bands are downconverted and time-space-decoded before the signal is sent into the building and to the UEs. Unfortunately, the mentioned time-space-coder and a time-space-decoder are not part of existing standards.
Furthermore, the design relies on outdoor units with signals in frequency bands that easily penetrate buildings, but may also cause interference around the building with outdoor mobile networks in the same frequency band. This concept also suffers from the so-called keyhole effect described in \cite{MBreiling_UE-SideVirtualMIMOUsing_2014} that leads to a MIMO channel rank reduction.

\subsection{Contribution and Outline}
In this paper, we describe a mmWave bridge as a simple solution that allows mobile network operators to fronthaul sub-6\,GHz small cells of any preferred vendor from existing outdoor cell sites to buildings or distributed short-range relays along a train track, using mmWave frequencies.
A mmWave bridge can be deployed to provide additional wireless capacity and coverage with a fixed wireless access (FWA) service if the cells on sub-6\,GHz frequency bands are already loaded and the allowed transmit power budget is exhausted, or the customer needs a dedicated mobile cell and it is too expensive or impossible to provide dedicated optical fiber connections for on-site small cells.
The analog signal up- and downconversion ensures transparent operation regarding the used radio access technology and readily integrates with existing base stations and smartphones, without new proprietary hardware or changes to wireless communication standards.
We built a mmWave bridge prototype to demonstrate the feasibility and to perform first tests.

The rest of this article is organized as follows: we first describe the deployment objectives and requirements and present the mmWave bridge system architecture.
This is followed by the description of our prototype implementation. 
Then, we report on measurements of the prototype as a proof of concept.
The last section concludes this work.

\section{mmWave Bridge Deployment and Architecture}
Our primary objective is to provide local coverage through small cells where it is impossible or too costly to deploy optical fiber connections.
Interference with existing cellular networks has to be avoided and their available capacity has to be maintained.
Specific deployment use-cases are described in the following section. Then an architecture overview is given, followed by functional aspects.

\subsection{Deployment Use-Cases}
\label{sec:deployment}
\begin{figure}
	\centering
	\includegraphics[width=\linewidth]{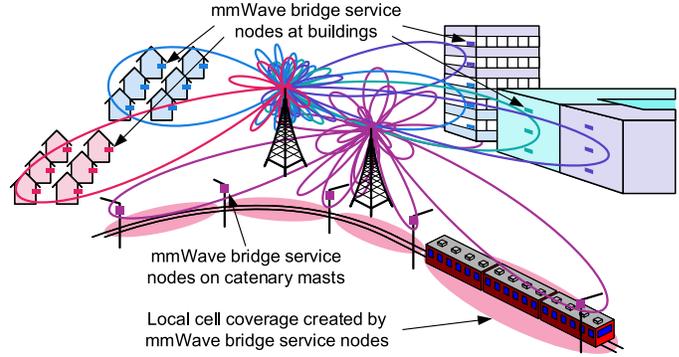}
	\vspace{-2em}
	\caption{Deployment examples for a mmWave bridge for FWA service to houses or large buildings and as a fronthaul for an RF corridor along a train track.\label{fig:mmwb_depl_ex}}
\end{figure}

A first use-case of the mmWave bridge is to provide FWA service to buildings into which the outdoor network poorly penetrates, see \fref{fig:mmwb_depl_ex}. A mmWave bridge donor node typically serves multiple buildings in its vicinity. The BS is shared among the users in the buildings covered by the donor link.
Depending on the environment and covering larger areas, multiple donor links concurrently serve groups of buildings.

A second use-case is to provide Internet connectivity and additional capacity to trains by means of an RF corridor along railway tracks realized with a distributed antenna system (DAS) \cite{Jamaly_DeliveringGigabitCapacitiesPassenger_2019}. The costly fiber fronthaul links and installation of the DAS are replaced with the mmWave bridge.
A donor node serves multiple mmWave bridge service nodes installed at regular distance intervals along the railway track to create the dedicated corridor cell, see \fref{fig:mmwb_depl_ex}.

\begin{figure*}
	\centering
	\includegraphics[width=\textwidth]{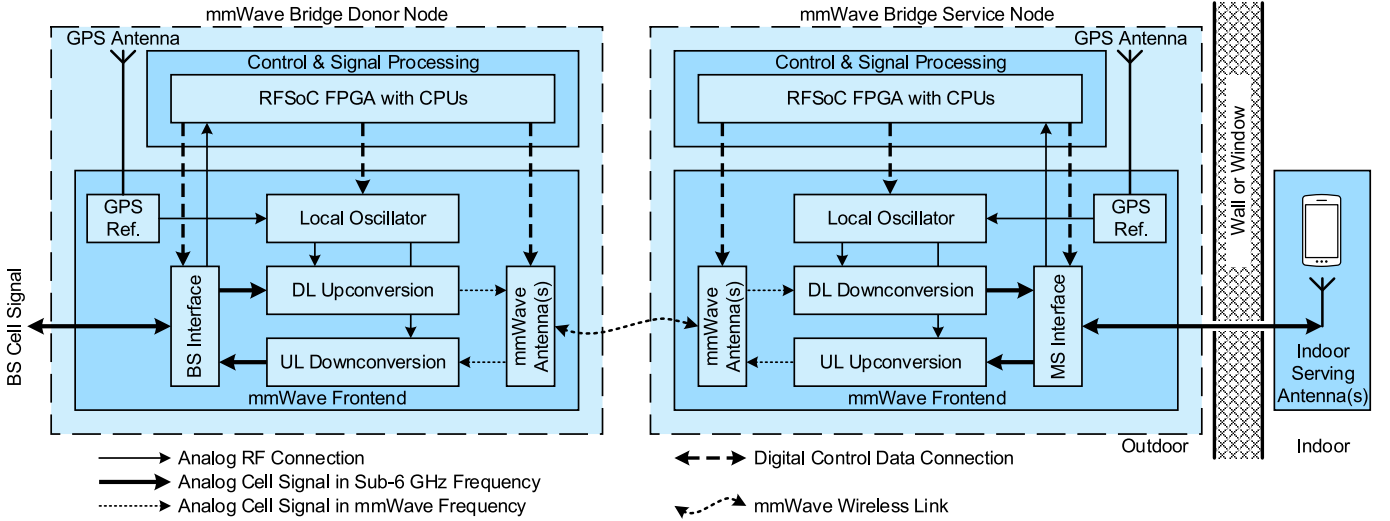}
	\vspace{-1.5em}
	\caption{Block diagram of the mmWave bridge donor node and service node. The serving antenna for the user access link is placed indoors.\label{fig:mmwb_sysdiag}}
	\vspace{-0.5em}
\end{figure*}

\subsection{Architecture Overview}
\label{sec:sysarch}
As a solution to the costly deployment of fiber and small cells, our proposed mmWave bridge allows installing the sub-6\,GHz small cell on existing cell sites where the fiber backhaul is available.
A donor node also installed at the cell site connects to the RF ports of the radio unit (RU) to replace the passive antennas of the small cell, see \fref{fig:mmwb_concept_example}.
The wireless access link of the small cell is relayed over the mmWave bridge between the donor node and the service node that is installed at, for example, a building where the small cell signal is emitted inside on the original sub-6\,GHz frequency.
One donor small cell connected to one mmWave bridge donor node serves one or multiple service nodes located in the coverage area of the donor node, allowing to share or dynamically move the small cell capacity between service nodes.

An overview of our proposed mmWave bridge is provided in \fref{fig:mmwb_sysdiag}.
The donor node and the service node both operate similarly: they upconvert the received sub-6\,GHz  signal to mmWave for retransmission over the donor link and downconvert the received mmWave signal to sub-6\,GHz for retransmission to the BS or UE.
To allow an immediate deployment and compatibility with wireless standard compliant small cells as well as mobile devices, our proposed mmWave bridge operates fully transparent to the RAT by keeping the cell signal in the analog domain.
For system management, a digital control and signal processing block is required.

\subsection{Beamforming}
To compensate for the higher path loss in mmWave frequencies while keeping the antenna dimensions compact and proportional to the wavelength, high-gain mmWave beamforming antennas are required.
Static scenarios allow for a fixed deployment in which mmWave beams are adjusted once during installation.
However, there are use-cases where the served mmWave bridge service nodes can vary. Examples are a cell following a train by always serving the mmWave bridge service nodes closest to the train, or FWA service where capacity can be moved between office buildings and residential buildings depending on, for example, the hourly demand or changes to the environment. In these cases, an adjustment to the beams between the donor and service nodes may be required. For such more agile deployments, an antenna with an electronically steerable beam is employed to ensure a stable link that minimizes the path loss between the nodes.

\subsection{Donor Link Multiple Access}
When more capacity needs to be deployed, for example, in densely built areas, multiple donor cells need to be relayed over mmWave bridges. The large available bandwidth at mmWave frequencies is used for multiplexing several sub-6\,GHz cells supporting frequency division multiple access (FDMA).
Additionally, mmWave beamforming antennas with a narrow beam pattern allow for multiplexing different cells in different directions providing spatial division multiple access (SDMA) for multiple small cells sharing the same mmWave frequency on their donor link.

\subsection{Duplexing Models}
Most cellular signals in the frequency bands below 3\,GHz operate in frequency division duplex (FDD), while cellular signals above 3\,GHz operate in time division duplex (TDD).
To use a single antenna for both transmission (TX) and reception (RX), different hardware configurations are required for FDD and TDD.
For FDD, duplexers \cite{Rappaport_WirelessCommunicationsPrinciplesPractice_1996} are used to separate TX and RX paths at the BS interface and mobile station (MS) interface (see \fref{fig:mmwb_sysdiag}).
Because corresponding bandpass filters are challenging to design at mmWave frequencies providing the required sharp passbands, separate mmWave antennas are used for downlink (DL) and uplink (UL) paths.
For TDD, solid-state RF switches \cite{Rappaport_WirelessCommunicationsPrinciplesPractice_1996} are used to switch the BS and MS interface and the mmWave antennas between TX and RX.

\section{Prototype Implementation}
\label{sec:implementation}
\begin{figure}
	\centering
	\includegraphics[width=\linewidth]{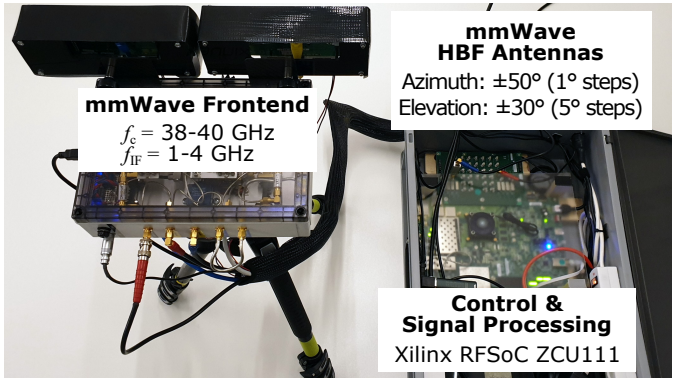}
	\vspace{-2em}
	\caption{Picture of one frontend with the mmWave HBF antennas and the control and signal processing board.\label{fig:photo_frontend}}
\end{figure}

Our initial prototype verifies the feasibility of relaying sub-6\,GHz standard-compliant cellular signals over mmWave frequencies and back to the original frequency to establish a successful connection with commercial off-the-shelf BS and UEs.
For simplicity, we currently support only SISO and FDD.
A picture of our mmWave bridge prototype is given in \fref{fig:photo_frontend} and the implementation of the components from \fref{fig:mmwb_sysdiag} is described in the following.

\subsection{mmWave Radio Frontend and Antenna}
\subsubsection{Up-/downconversion and Filtering}
The mmWave bridge frontends for the RF signal up- and downconversion are implemented using commercial off-the-shelf components.
Cavity filter diplexers are used for interfacing the mmWave bridge frontends with the base station on the donor node and the service antenna on the service node.
For the upconversion, the sub-6\,GHz signal is mixed with a continuous wave local oscillator (LO) signal from a configurable signal generator.
To prevent the carrier frequencies from drifting, the LOs in both nodes are fed with a stabilized reference signal from Trimble \mbox{Thunderbolt-E} Global Positioning System (GPS) disciplined oscillators.
A bandpass filter with a passband from 38--40\,GHz suppresses the unwanted sideband right after the mixer. Finally, the upconverted signal is amplified with a power amplifier and fed to the transmitting mmWave antenna.

In the downconversion path, the received signal is amplified with a low-noise amplifier (LNA).
After the LNA, a bandpass filter ensures that only signals in the range 38--40\,GHz are mixed to the original sub-6\,GHz carrier frequency.
A bandpass filter at the output of the downconverter finally passes the desired signal to the diplexer for the base station on the donor node or the serving antenna on the service node.

\subsubsection{mmWave Antennas}
In our prototype, we use separate mmWave antennas for DL and UL.
On the service node, we can use open waveguides with 6\,dBi gain or horn antennas with 20\,dBi gain.
Two electronically steerable beamforming antennas are used on the mmWave donor node.
We use Holographic Beam Forming (HBF) \cite{EricJBlack_HolographicBeamformingMIMO_2017} antennas from Pivotal Commware, which allow significant flexibility and speed in terms of beam steering. These antennas provide a gain of around 20\,dBi.
The single-polarization analog beamforming antennas are controlled over a USB interface.

\subsection{Control and Signal Processing}
\label{sec:rfsoc}
For analyzing the DL signal and controlling the beamforming antenna, the Xilinx RFSoC evaluation board ZCU111 was chosen.
This board provides RF input and output ports, an Ethernet port connected to a mobile router for remote access, a USB port, and two integrated processors. The larger quad-core processor runs a Petalinux, and with PYNQ \cite{Stewart_SoftwareDefinedRadioRFSoC_2018}, it provides the possibility to use Python Jupyter notebooks for high-level control tasks.
At the donor node, the received and downconverted DL signal is split and fed to the field programmable gate array (FPGA) to be processed for the beam control.
Upon initialization of the mmWave link, a beam control routine is run. The service node is configured to send a pilot signal while the donor node rapidly scans all configured beams. The receive and transmit antennas of the donor node are then configured to the beam providing the strongest signal. This procedure can be triggered periodically by sacrificing a short sub-6\,GHz link interruption below the RAT connection reestablishment time.

\section{Experimental Results}
\label{sec:results}
To validate the mmWave bridge concept and test our prototype, we conducted several measurements and report representative examples in the following.

For a first measurement for functional verification, we used a commercial base station providing a 4G Long Term Evolution (LTE) Release~15 FDD cell with 20\,MHz of bandwidth. One antenna port of the RU was connected to the donor node of our mmWave bridge SISO prototype.
The mmWave bridge service node was set-up in the lab with line-of-sight (LOS) to the donor node and the MS interface was cabled to a small RF shielded Faraday box with an antenna and UE inside.
System parameters were measured on the DL signal from the mmWave bridge service node fed to the antenna in the Faraday box.
A Rohde~\& Schwarz FSH8 spectrum analyzer with LTE downlink signal analysis functionality was used to measure reference signal received power (RSRP), carrier frequency offset, overall error vector magnitude (EVM), and signal-to-noise ratio (SNR).
Additionally, data throughput tests were performed with iPerf3 and five parallel connections.
With a DL RSRP of \textminus44.8\,dBm fed to the Faraday box antenna, an overall EVM of 4.6 percent and an SNR of 34.5\,dB were measured. Two LTE category~18 and 20 UEs achieved a median throughput of 97.8\,Mb/s and 96.7\,Mb/s, respectively. This is close to the theoretical maximum data rate of 98\,Mb/s.

\begin{figure}
	\centering
	\includegraphics[width=\linewidth]{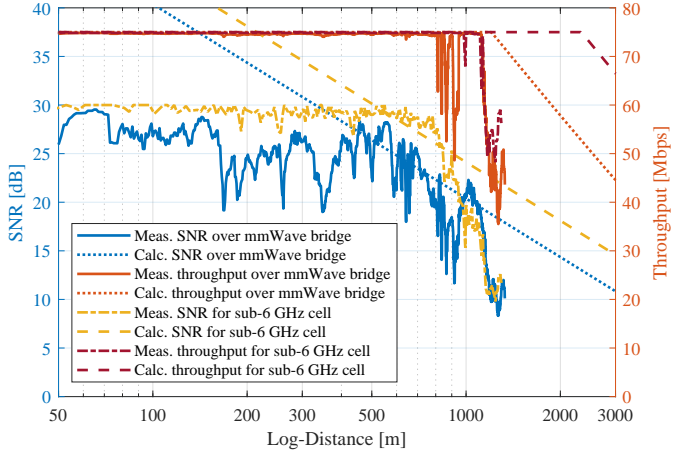}
	\vspace{-2em}
	\caption{Measurement results for 50\,m up to 1300\,m and calculated results up to 3000\,m. Solid lines represent measured parameters from a UE when using the mmWave bridge with 30\,dBm EIRP, dash-dot lines represent the measured parameters from a UE directly communicating with the sub-6\,GHz cell also with 30\,dBm EIRP. Dotted and dashed lines represent calculated results for the mmWave bridge and direct sub-6\,GHz cell, respectively.\label{fig:meas_comb_plot}}
\end{figure}

For a second measurement, a compact LTE Release~10 base station with integrated evolved packet core (Nutaq PicoLTE) was used. This allows a peak data rate of 75\,Mb/s for a SISO link with 20\,MHz of bandwidth in the 2.6\,GHz frequency band.
The base station RF signal was cabled to the mmWave bridge donor node with an antenna height of 2\,m and installed in a rural area on a straight track with a length of 1.3\,km. The mmWave bridge service node with an antenna height of 1.75\,m was cabled to the RF shielded Faraday box with a UE inside, as in the first measurement. With a drive-test software, we logged various parameters from a smartphone, such as the SNR and throughput, while moving the mmWave bridge service node away from the donor node, always in LOS.
The measured parameters are filtered with a moving average filter over 5\,m, and the SNR and throughput are shown with solid lines in \fref{fig:meas_comb_plot}. Note that UEs often do not report SNR values $>$30\,dB, therefore we can see a saturation effect.
Measurements are available for a distance of 50\,m up to 1300\,m. For validation and extrapolation beyond 1300\,m, the SNR is calculated based on the free space path loss model and the estimated noise floor from the measurements.
Using this SNR, the throughput is calculated with the throughput model from \cite[Sec.~A.2]{zz3gpp.36.942}. Extrapolated SNR and throughput are shown with dotted lines in \fref{fig:meas_comb_plot}.
For comparison, the measured SNR and throughput of the sub-6\,GHz cell with the same transmit power are shown with dash-dot lines, and the respective calculated parameters are shown with dashed lines in \fref{fig:meas_comb_plot} for validation.
We used a transmit power of 30\,dBm effective isotropic radiated power (EIRP) for the mmWave bridge and the sub-6\,GHz cell.

The measurements show that the displaced cell maintains its maximum throughput for a mmWave bridge donor link distance of up to 800\,m. This is comparable to a direct sub-6\,GHz connection, which maintains its maximum throughput up to 1100\,m due to the lower propagation attenuation.

Regarding the FWA use-case, we consider a 20\,dB building entry loss resulting in a required sub-6\,GHz transmit power of 50\,dBm EIRP to reach the SNR and throughput shown in \fref{fig:meas_comb_plot} inside a building at the given distance.
Because the mmWave bridge service node is installed with the mmWave donor link antennas outside and the sub-6\,GHz service link antennas inside, the mmWave bridge avoids the need to boost the outside cell power to compensate for the building entry loss.
By creating a displaced and isolated cell inside the building without occupying valuable sub-6\,GHz spectrum outside for the connection, a reduction in the outdoor network capacity is avoided.

Considering the use-case of an RF corridor along a train track, the mmWave bridge indeed allows providing a sub-6\,GHz cell signal with an SNR $>$20\,dB in the surrounding of the service node that is up to 800\,m away from the donor node without the need for fiber deployment along the track and with a mmWave EIRP of only 30\,dBm. This range can be extended by increasing the transmit power of the bridge without disturbing the cellular network on sub-6\,GHz frequencies along the track.
The use of mmWave frequencies for the out-of-band signal relaying allows for compact high-gain antennas and a less complex and costly implementation compared to an in-band full-duplex repeater, often requiring intricate and expensive self-interference suppression.

\section{Conclusion}
\label{sec:conclusion}
Providing high data capacities indoors is challenging if optical fiber is not available, and the outdoor cellular network is already loaded. Spectrum in higher frequency ranges allows for wider signal bandwidth and larger capacities; however, these signals do not (or not easily) penetrate buildings.
We propose a mmWave bridge that transfers sub-6\,GHz cell signals from a base station site to buildings or along train tracks using mmWave frequencies and provides the original sub-6\,GHz cell inside buildings or to the train.
This has several advantages.
First, the mmWave bridge operates transparent to the cellular signals, readily working with existing base stations and user equipment.
Second, by using the newly available mmWave spectrum, no interference to existing outdoor cellular networks is added, nor are any changes in the existing network necessary.
Third, using the large available bandwidth in mmWave spectrum for frequency division multiple access and directive mmWave beamforming antennas for spatial division multiple access, many cells can be multiplexed for a high area capacity.

A prototype has been built consisting of analog frontends for signal up- and downconversion, mmWave beamforming antennas, and a control and signal processing board for estimating various signal parameters, setting the local oscillator frequencies, and steering the mmWave beam.
Most importantly, with this prototype, we can show that the mmWave bridge concept works.
Measurement results show that smartphones can achieve their peak throughput even at donor link distances of up to 800\,m when using only 30\,dBm EIRP for a 20\,MHz cell signal.

Concluding, the mmWave bridge has the potential to complement a cellular network by providing a wireless fronthaul link for delivering increased data capacities inside buildings and along train tracks.

\section*{Acknowledgment}
The authors would like to thank Pivotal Commware Inc., Kirkland (WA), USA for providing the mmWave antennas, D.~Northcote from the University of Strathclyde, Glasgow, UK for support around PYNQ on the Xilinx RFSoC, R.~Ghanaatian for implementing the FPGA framework, and E.~Zimmermann for implementing the frontends.

\bibliographystyle{IEEEtran}
\bibliography{references}

\vspace{-2\baselineskip}
\begin{IEEEbiographynophoto}{\textsc{Adrian Schumacher}} [M'07,SM'21]
received the M.Sc. degree from the Royal Institute of Technology (KTH), Stockholm, Sweden, in 2006. He is a senior engineer at Swisscom and currently working towards a Ph.D. degree at the École Polytechnique Fédérale de Lausanne (EPFL), Lausanne, Switzerland.
His research interests include digital signal processing, wireless systems, and mmWave communications.
\end{IEEEbiographynophoto}

\vspace{-2\baselineskip}
\begin{IEEEbiographynophoto}{\textsc{Ruben Merz}}
earned the Ph.D. and M.Sc. degrees from the School of Computer and Communication Systems at EPFL, and a Master of Advanced Studies degree in management, technology and economics at ETH Z\"urich.
He is the lead system architect for the 5G program at Swisscom. Before joining Swisscom, Ruben was a Senior Scientist at Telekom Innovation Laboratories, in Berlin, Germany.
\end{IEEEbiographynophoto}

\vspace{-2\baselineskip}
\begin{IEEEbiographynophoto}{\textsc{Andreas P. Burg}}
(Member, IEEE) received the Dipl.Ing. degree from the Swiss Federal Institute of Technology (ETH), Zurich, Switzerland, in 2000, and the Dr.Sc.Techn. degree from the Integrated Systems Laboratory, ETH Zurich, in 2006.
In 2011, he became a Tenure Track Assistant Professor with the École Polytechnique Fédérale de Lausanne (EPFL), where he is currently leading the Telecommunications Circuits Laboratory. He was promoted to Associate Professor in June 2018.
\end{IEEEbiographynophoto}

\vfill
\end{document}